\documentclass[aps,prl,showpacs,superscriptaddress,floatfix,twocolumn]{revtex4-1}
\usepackage{times}
\usepackage{graphicx}
\usepackage{amsmath}
\usepackage{amssymb}
\usepackage{subfigure}
\usepackage{color}
\usepackage{epstopdf}
\usepackage{textcomp}
\usepackage{gensymb}

\begin{document}

\title{Electronic Correlation Induced Expansion of Compensated Electron and Hole  Fermi Pockets in $\delta$-Plutonium}

\author{Roxanne Tutchton}
\email{rtutchton@lanl.gov}
\affiliation{Theoretical Division, Los Alamos National Laboratory, Los Alamos, New Mexico 87545, USA}

\author{ Wei-ting Chiu}
\affiliation{Department of Physics, University of California,  Davis, California 95616, USA}

\author{R. C. Albers}
\affiliation{Theoretical Division, Los Alamos National Laboratory, Los Alamos, New Mexico 87545, USA}

\author{G. Kotliar} 
\affiliation{Department of Physics and Astronomy, Rutgers University, Piscataway, New Jersey 08854, USA}

\author{Jian-Xin Zhu}
\email{jxzhu@lanl.gov}
\affiliation{Theoretical Division, Los Alamos National Laboratory,
Los Alamos, New Mexico 87545, USA}
\affiliation{Center for Integrated Nanotechnologies, Los Alamos National Laboratory,
Los Alamos, New Mexico 87545, USA}

%$^{\ast}$ e-mail: jxzhu@lanl.gov
%$^{\ast}$ These authors contributed equally to this work

%
%\author{Jian-Xin Zhu}
%\email[To whom correspondence should be addressed. \\ Electronic address: ]{jxzhu@lanl.gov}
%\homepage{http://theory.lanl.gov}
%\affiliation{Los Alamos National Laboratory,
%Los Alamos, New Mexico 87545, USA}
%
%\author{R. C. Albers}
%\affiliation{Los Alamos National Laboratory,
%Los Alamos, New Mexico 87545, USA}
%
%\author{K. Haule}
%\affiliation{Rutgers University, Piscataway, New Jersey 08854, USA}
%
%\author{G. Kotliar}
%\affiliation{Rutgers University, Piscataway, New Jersey 08854, USA}
%
%\author{J. M. Wills}
%\affiliation{Los Alamos National Laboratory,
%Los Alamos, New Mexico 87545, USA}

%\begin{abstract}

\begin{abstract}
%The Fermi surface is a fundamental electronic property of all solids. In this paper we provide, for the first time, detailed calculations of the Fermi surface and associated mass renormalizations of the delta phase of Plutonium ($\delta$-Pu). 
Plutonium is a critically important material as the behavior of its 5{\it f}-electrons stands midway between the metallic-like itinerant character of the light actinides and localized atomic-core-like character of the heavy actinides. The $\delta$-phase of plutonium ($\delta$-Pu), while still itinerant, has a large coherent Kondo peak and strong electronic correlations coming from its near-localized character. Using sophisticated Gutwiller wavefunction and dynamical mean-field theory correlated theories, we study for the first time the Fermi surface and associated mass renormalizations of $\delta$-Pu together with calculations of the de Haas-van Alphen (dHvA) frequencies. We find a large ($\sim$200\%) correlation-induced volume expansion in both the hole and electron pockets of the Fermi surface in addition to an intermediate mass enhancement. All of the correlated electron theories predict, approximately, the same hole pocket placement in the Brillouin zone, which is different from that obtained in conventional density-functional band-structure theory, whereas the electron pockets from all theories are in, roughly, the same place.
\end{abstract}
 %\end{abstract}
%\pacs{74.25.Jb, 74.20.Pq, 71.27.+a, 71.28.+d}
%74.25.Jb	Electronic structure (photoemission, etc.)
%71.27.+a	Strongly correlated electron systems; heavy fermions
%74.20.Pq	Electronic structure calculations (for methods of electronic structure calculations, see 71.15.-m)
%71.28.+d	Narrow-band systems; intermediate-valence solids

\maketitle

\newpage

\begin{figure*}[htb]
\begin{center}
\includegraphics[width=0.8\textwidth]{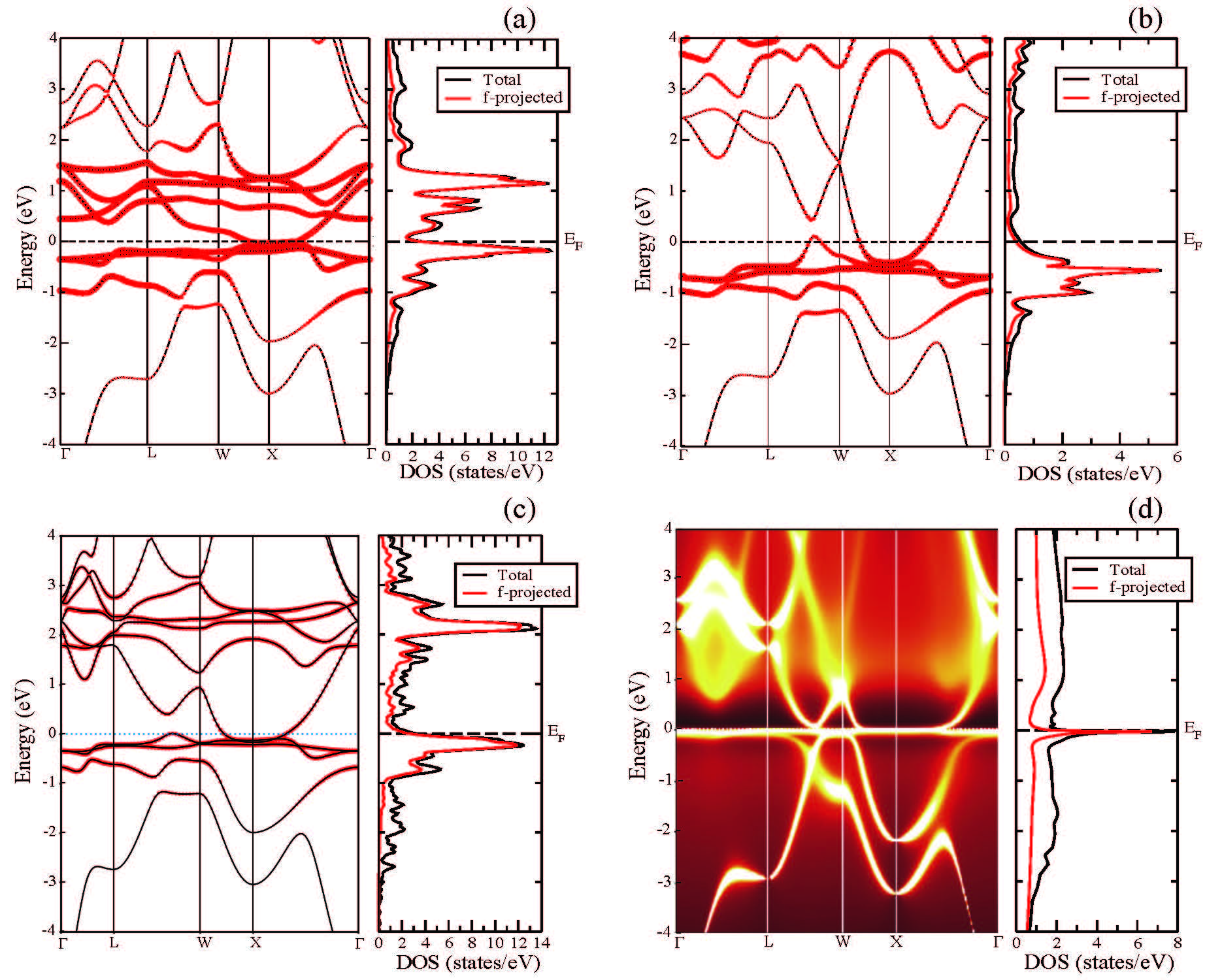}
\end{center}
\caption{ {\bf The electronic band structures} and densities of state (DOS) for (a) non-magnetic $\delta$-Pu calculated using the GGA with no explicit electron-electron correlation, (b) non-magnetic $\delta$-Pu calculated with constrained GGA+{\it U} for $U=4.5$ eV and $J=0.512$ eV, (c) paramagnetic  $\delta$-Pu calculated using GGA+GutzA for $U=4.5$ eV and $J=0.512$ eV, and (d) paramagnetic $\delta$-Pu calculated using  GGA+DMFT for $U=4.5$ eV and $J=0.512$ eV. The thick red bands indicate the {\it f}-electron occupations, and the red curves on the DOS show the {\it f}-projected densities. The path that the band dispersions take through the reciprocal space is shown in Fig.~\ref{BZsym}. Note that while the band-structure scales are all the same (between $-4$ and $+4$ eV) that the DOS scales have different maximums, which should be taken into account when comparing the effects of correlation on the DOS of the four different methods. We have changed the scales to best show the structure in the DOS of each method.} 
\label{El-struct}
\end{figure*}

\begin{figure*}[htb]
\begin{center}
\includegraphics[width=0.8\textwidth]{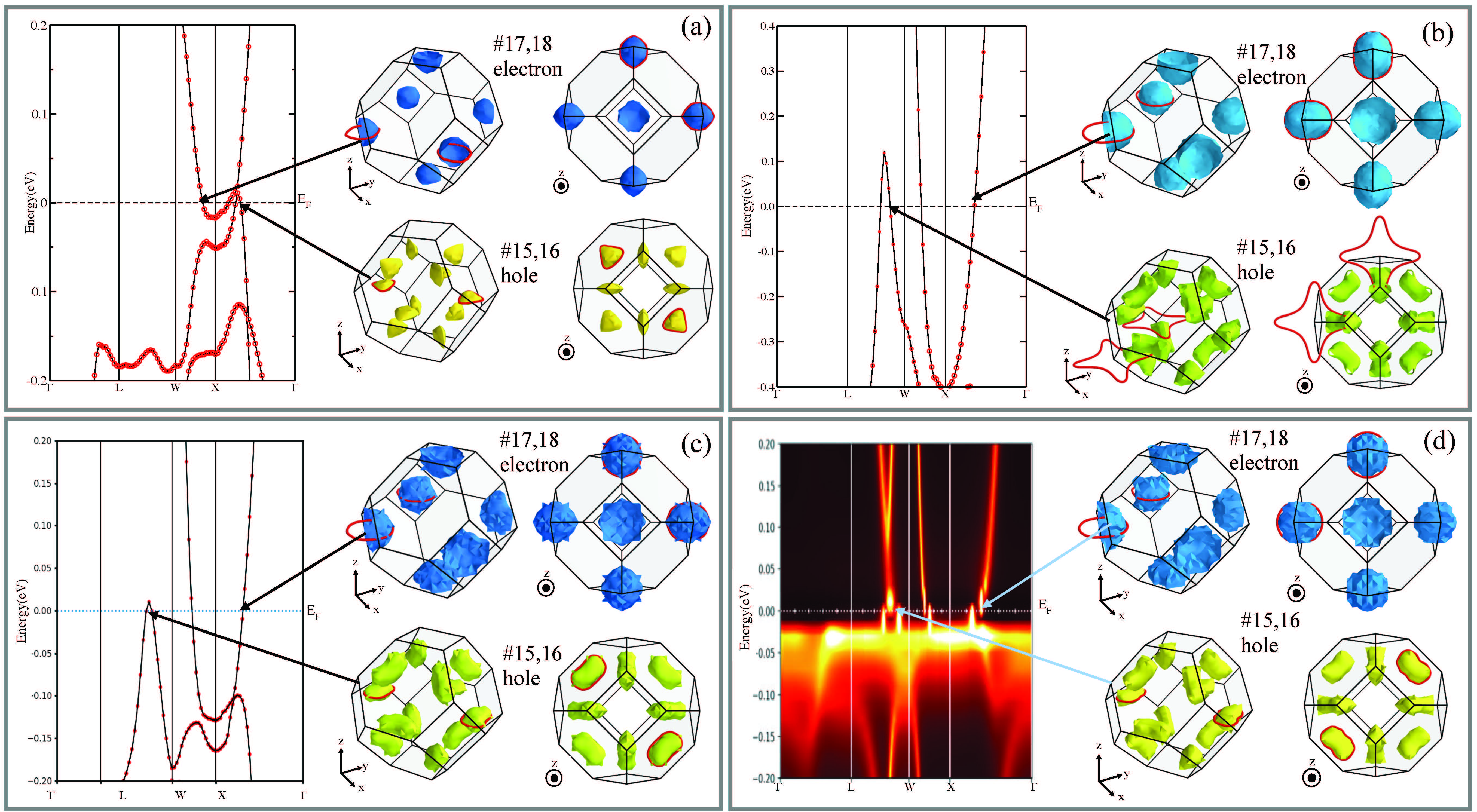}
\end{center}
\caption{ {\bf The Fermi surface topologies} and corresponding, close-up band dispersions for (a) the GGA method, (b) the GGA+{\it U} method, (c) the GGA+GutzA methods, and (d) the GGA+DMFT method. Each panel shows the hole (bands 15 and 16) and electron (bands 17 and 18) isosurfaces for two orientations as indicated by the coordinates to the bottom left of each Brillouin zone (BZ). The arrows from the Fermi surfaces to the band structures indicate the band intersections with the Fermi energy that correspond to each Fermi surface. Red lines are interposed on the Fermi surfaces to indicate the cross-sections related to the dHvA extremal frequencies for an external {\bf B} field parallel to the $z$-axis (${\bf B}\parallel z$). The path the band dispersions take through the BZ is shown in Fig.~\ref{BZsym}.} 
\label{ElZoom_FStop}
\end{figure*}

\begin{figure}[htb]
\begin{center}
\includegraphics[width=0.9\columnwidth]{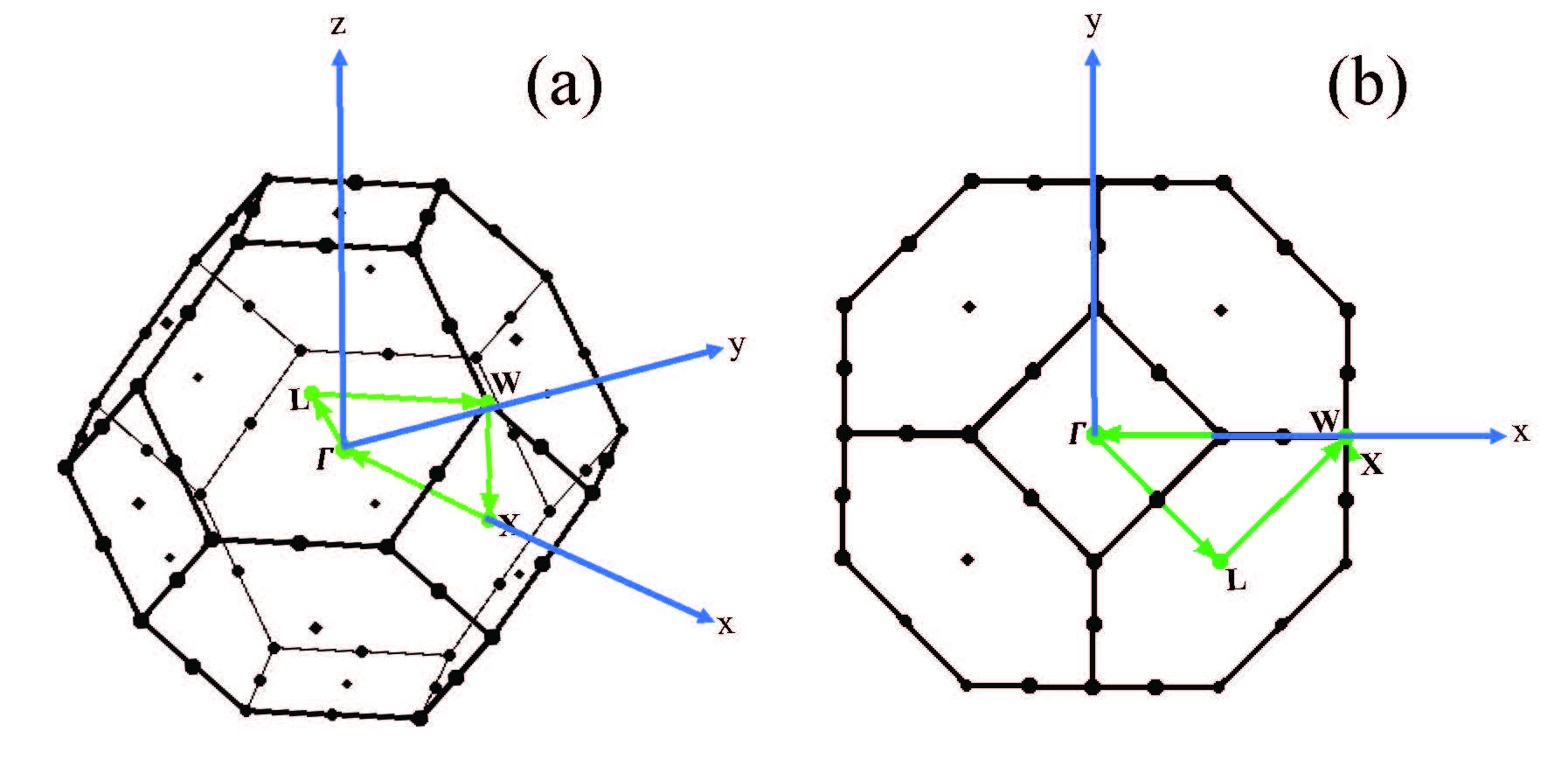}
\end{center}
\caption{ {\bf The high symmetry path through the first Brillouin zone} used for the band structure calculations goes from $\Gamma \rightarrow L\rightarrow W\rightarrow X\rightarrow \Gamma$.} 
\label{BZsym}
\end{figure}

%{\it Introduction:} 
With six allotropic crystal phases at ambient pressure, Pu is one of the most complex elemental solids in the periodic table. Much of its exotic behavior is driven by a large anisotropy natural in its {\it f}-electron bonding whose strength is, in turn, tuned by strong electronic correlations, which arises from its 5{\it f}-electron behavior standing midway between the metallic-like itinerant character of the light actinides and the localized atomic-core-like character of the heavy actinides. In addition these correlations are highly temperature dependent and give rise to a strong atomic volume dependence of the different phases. In particular, the 25\% volume expansion between the $\alpha$ and $\delta$ phases gives rise to exotic physics that has appealed to experimentalists and theorists for decades~\cite{ 1985Weinberger, 2001Savrasov, 2002Terry, 2003Dai, 2006Kotliar, 2015Soderlind, 2013Zhu, 2019Soderlind,2019Brito}. Consequently, the {\it f}-electrons in $\delta$-Pu appear to exhibit a complex combination of localized and itinerant characteristics making the local moments and their impact on the electronic structure difficult to model~\cite{1996MeotReymond, 2001Savrasov, 2012Zhu, 2015Janoschek}.\\
\indent Associated with this 5$f$-electron duality, the magnetic behavior in Pu has long been a puzzle in the condensed matter and materials physics community~\cite{2006Heffner, 2005Lashley, 2017Janoschek, 2017Migliori} since strongly localized electrons typically exhibit some form of magnetism. No long-range magnetic ordering has been observed for any phase of Pu~\cite{2006Kotliar, 2015Soderlind, 2017Janoschek, 2019Soderlind}. However, theoretical calculations performed by using density functional theory (DFT), within either the local density approximation (LDA) or generalized gradient approximation (GGA), cannot reproduce the characteristic volume expansion for $\delta$-Pu unless  spin and/or orbital polarization are included~\cite{1991Solovyev, 1994Soderlind, 2000Bouchet, 2000Savrasov, 2001Soderlind, 2002Soderlind, 2003Kutepov, 2019Soderlind}. The same is true for DFT with the addition of a Coulomb parameter $U$ (LDA+{\it U} and GGA+{\it U}). It has long been believed that the level of electronic correlation in these approaches were inadequate, and, indeed, strongly-correlated electron theoretical methods have recently shown very promising results. By modeling the competition between the on-site Coulomb repulsion among localized $f$-electrons and their hybridization with the  itinerant electrons in a quantum impurity fashion, dynamical mean-field theory (DMFT) is able to describe the single-particle excitation properties of $\delta$-Pu in good agreement with photoemission spectroscopy measurements~\cite{2001Savrasov, 2002Terry, 2007Zhu} as well as to predict the valence fluctuations validated by inelastic neutron spectroscopy~\cite{2007Shim, 2015Janoschek}. Another strongly-correlated electron theoretical approach uses the Gutzwiller wavefunction approximation (GutzA)~\cite{1965Gutzwiller, 2015Lanata, 2017Lanata}, which has been used to successfully calculate the volume dependence in Pu phases without the addition of artificial orbital polarization~\cite{2006Julien, 2015Lanata}. The combination of LDA (or GGA) and GutzA has the same mathematical structure as LDA combined with DMFT~\cite{2006Kotliar}, with the difference that the GutzA assumes infinite quasiparticle lifetimes, which is equivalent to neglecting the incoherent component of the electrons. This makes the GutzA method less accurate for calculating strong correlation effects than DMFT. However it has the important compensating advantage of being significantly less computationally demanding~\cite{2015Lanata}.\\
\indent Recently, an attempt has been made to bring magnetism back into the models with the ``disordered local moment”~\cite{2003Niklasson}, involving a static DFT model. The contrasting picture is the ``valence fluctuation''~\cite{2015Janoschek} (dynamical) model using the DMFT approach described above. The static model describes the orbitals as individual localized magnetic moments that are spatially and temporally disordered such that any long-range magnetic ordering is obscured by averaging over time. The dynamical model instead proposes a quantum entanglement between the localized magnetic moments and the itinerant conduction electrons, resulting in valence fluctuations that effectively screen the magnetic ordering below the Kondo temperature, $T_{K}$, of the material~\cite{2015Janoschek}.\\
\indent In an effort to determine the validity of these theoretical models, previous studies of Pu allotropes have focused on reproducing the volume expansion and bulk modulus of $\delta$-Pu~\cite{2000Jones, 2000Nordstrom, 2001Savrasov, 2003Wills, 2005Wu, 2006Julien, 2008Soderlind, 2015Lanata, 2019Soderlind}. This work strives to fill a gap in the understanding of the momentum space electronic behavior through a direct examination of strong correlation effects. We accomplish this by probing the Fermi surface topology of $\delta$-Pu using the de Haas-van Alphen (dHvA) effect present in metallic systems~\cite{1982Shoenberg, 2012Rourke, 2013Wang}, and comparing results for several theoretical methods including DFT within the GGA, GGA+{\it U} (where {\it U} is the Hubbard parameter defining the strength of a static Coulomb interaction), the Gutzwiller approximation (GGA+GutzA), and dynamical mean-field theory (GGA+DMFT). The most striking result includes the discovery of strong correlation induced expansion of compensated electron and hole pockets as well as the relocation of hole pockets. \\
\\
\noindent
{\bf Results and Discussion}
\\
{\bf Electronic Structure.} The first of the theoretical methods we explored was the standard DFT GGA calculation, which does not explicitly incorporate electronic correlation beyond a simple local exchange-correlation function. The electronic band structure and density of states (DOS) calculated using this method are shown in Fig.~\ref{El-struct}(a), and the Fermi surface calculations are given in Fig.~\ref{ElZoom_FStop}(a). This calculation is the baseline for the other three methods, each of which adds additional correlation effects using increasingly sophisticated techniques. The GGA+{\it U} method applies a static Coulomb interaction to the system within the Kohn-Sham formalism. The GGA+GutzA uses an auxiliary particle theory, while assuming infinite quasiparticle lifetimes. The GGA+DMFT is the most sophisticated and computationally demanding method for calculating electronic structure with strong correlation effects. The resulting band structure for each method is shown in Fig.~\ref{El-struct} along with its accompanying density of state (DOS). It should be noted that the structure in the DOS for GGA+DMFT is greatly washed out relative to the other three methods. This is due to quasiparticle lifetime effects naturally included in GGA+DMFT, whereas the other three methods have sharp (infinite lifetime) quasiparticle states.\\
\indent The effects of electronic correlation on the quasiparticle band structure is to shift and/or renormalize (distort) the energy bands. The addition of a static Coulomb parameter with GGA+{\it U}  (Fig.1(b)) results in a large shift of the conduction bands just above the Fermi energy and a distortion of the upper valence bands with respect to the GGA results (Fig.~\ref{El-struct}(a)). The GGA+{\it U}  DOS also shows a distinct splitting of the 5$f$-electron subshells, $j=5/2$ and $j=7/2$, with the $j=5/2$ electron character mainly just below the Fermi energy and the $j=7/2$ electron character between about 4 to 6 eV for the value of U we have chosen. The DOS calculated with GGA+GutzA shows a similar but less dramatic splitting of 5{\it f} subshells (in part due to spin-orbit coupling) (Fig.~\ref{El-struct}c). The corresponding band dispersion indicates a concentration of 5{\it f}-electrons from the $j=5/2$ sub-shell around the Fermi energy. There is also an upward shift of the conduction bands and accompanying renormalizations around the Fermi energy as compared to the GGA calculation of Fig.~\ref{El-struct}(a). This results in flattening of the bands around the $X$ point with increased slopes to the right and left of the flat region. The flattening of the bands in this region is more pronounced than the similar feature in the GGA+{\it U} band dispersion (Fig.~\ref{El-struct}(b)), while the shift and resulting slope increase is much more prevalent in the GGA+{\it U} method. The band dispersions calculated using GGA+DMFT (Fig.~\ref{El-struct}(d)) show by far the largest band renormalization of {\it f}-electrons around the Fermi energy, which is evident from the peak in the DOS (the Kondo peak). This feature is associated with the flattening of the top valence bands and bottom conduction bands around the Fermi energy. Comparing the GGA+GutzA and GGA+DMFT to the GGA+{\it U}  band structures, we find that the static approach of the GGA+{\it U} method does not include the type of band flattening or renormalization of the {\it f} bands expected from strong electronic correlations. \\
\\
{\bf Fermi surface topology.} We have further explored the behavior of the electronic structure around the Fermi energy is through an analysis of the Fermi surfaces shown in Fig.~\ref{ElZoom_FStop} along with sections of the band structure around the Fermi energy. The Brillouin Zone and high-symmetry paths used in Fig.~\ref{El-struct} is displayed in Fig.~\ref{BZsym}. At first glance, there is a qualitative similarity between the Fermi surfaces from each of the calculation methods. A notable difference in the GGA results is the location of the hole bands (\#15,16) intersection with the Fermi energy. The band structures that include some form of correlation effect show the hole pocket occurring between the $L$ and $W$ high-symmetry points, whereas the GGA method has the hole pocket between the $X$ and $\Gamma$ points. The rough position of the electron pockets seems unchanged between GGA and more correlated electron methods.\\
\indent We also find differences in dHvA frequencies, $f_{i}$, which are related to the area of the Fermi surface by the expression $f_{i}=1/\Delta(\frac{1}{{\bf B}})=\hbar A_{i}/2 \pi e$, where $e$ is the elementary charge of an electron, and $A_{i} $ is the extremal cross-sectional area of the $i^{th}$ branch of the Fermi surface in a plane perpendicular to the applied magnetic field, {\bf B}. These frequencies are denoted by the red lines around the isosurfaces in Fig.~\ref{ElZoom_FStop} , as well as listed in Table~\ref{dHvAT0}. In this case, we have calculated the extremal frequencies using a simulated external magnetic field parallel to the Cartesian $z$-axis as indicated in Fig.~\ref{BZsym}. A full analysis of the angular dependence on the dHvA frequencies and cyclotron masses is available in the Supplementary Information (SI) document.\\ 
\begin{table*}
\caption{{\bf The dHvA and volume data for Fermi surface calculations at zero temperature.} Frequencies are given in kilotesla (kT), and corresponding effective masses are in units of electron mass ($m_{e}$). Reciprocal occupied (electron bands 17 and 18) and unoccupied (hole bands 15 and 16)  Fermi surface volumes are given in units of $\AA^{-3}$. Calculations for $\delta$-Pu done with DFT, GGA+{\it U}, GGA+GutzA, and GGA+DMFT are compared. All measurements were taken for a magnetic field parallel to the $z$ axis (${\bf B}\parallel z$). The results for $T=0$ K were extrapolated from the GGA+DMFT calculation at $T=116$ K, where the sensitivity of electronic self-energy to temperature is negligible since this temperature is already far below the coherence temperature~\cite{2007Shim}. The Fermi energy was shifted by $-3.36$ meV to maintain the conservation of the number of electrons as $T\rightarrow0$ K.
\label{dHvAT0}}
\begin{center}
\begin{tabular}{| l | r | r | r | r | r | r | r | r | r | r | r | r |}
\hline
&\multicolumn{3}{c|}{DFT} &\multicolumn{3}{c|}{GGA+{\it U}} &\multicolumn{3}{c|}{GGA+GutzA} & \multicolumn{3}{c|}{GGA+DMFT} \\
\hline
Band & $f$ & $m^{*}$ & $V_{FS}$ & $f$ & $m^{*}$ & $V_{FS}$ & $f$ & $m^{*}$ & $V_{FS}$ & $f$ & $m^{*}$ & $V_{FS}$\\ 
\hline
15, 16 & $1.74$ & $3.19$ & $0.31$ & $10.2$ & $2.35$ & $1.02$ & $3.12$ & $2.00$ &  $0.91$ & $3.07$ & $1.84$ & $0.95$ \\ 
\hline
17, 18 & $2.87$ & $5.32$ & $0.31$ & $6.65$ & $1.54$ & $1.02$ & $6.27$ & $2.48$ & $0.91$ & $6.18$ & $2.17$  & $0.95$ \\
\hline
\end{tabular}
\end{center}
\end{table*}
\indent The most apparent development due to correlation effects is the expansion of both the electron and hole Fermi surfaces. This can be seen in the increase of the dHvA frequencies, but expansion is clearest in the reciprocal volumes given in Table~\ref{dHvAT0}. Compared to the GGA method, the GGA+{\it U}, GGA+GutzA, and GGA+DMFT reciprocal volumes increase by between ~200\% and ~230\%, with the largest increase occurring between the GGA and GGA+{\it U} methods. The nature of this volume expansion is consistent with the Luttinger theorem~\cite{1993Hewson}, which requires that the number of electrons be conserved and is directly proportional to the volume of the Fermi surface. Given that there are four bands (two degenerate pairs: 15\&16 and 17\&18) intersecting with the Fermi energy, and assuming the lower energy valence bands do not interact, as the electron pocket increases in volume the hole pocket must increase to compensate, thereby conserving the total electron number. This leads to the simultaneous expansion of the Fermi surfaces of both the hole and electron bands. This observation of Fermi pocket expansion is striking and unique for our multi-band system, given that the strong-correlation effect does not change the Fermi surface topology in a single-band correlated electron model~\cite{1993Hewson}.\\
\indent We have also analyzed the cyclotron effective masses on the Fermi surface. Their relationship to the dHvA cross-sectional can be expressed as $m^{*} = \frac{\hbar^{2}}{2\pi e}\left. \frac{\partial A}{\partial E}\right\rvert_{E=E_{F}}$, which is in units of the electron mass, $m_{e}$. The results of these calculations are recorded in Table~\ref{dHvAT0}. They show a decrease in the effective mass for methods including electronic correlations. As the cyclotron mass is calculated directly at the Fermi energy, we can explain the decrease in effective mass as stemming from the conspiring Fermi surface expansion and band renormalization. Because the effective mass is inversely proportional to the slope of the energy band, a steeper band intersecting the Fermi energy leads to a smaller cyclotron mass. From the sections of the electronic band structures shown in Fig.~\ref{ElZoom_FStop} it is apparent that the slope of the hole and electron bands increases when moving from GGA to the other 
strongly-correlated electron methods. The electron bands (\#17,18) calculated using GGA+{\it U} have the steepest slope and, consequently, the most reduced effective mass. Similarly, the band structures from the GGA+GutzA and GGA+DMFT show bands of increased slope at the Fermi energy resulting in smaller effective masses.\\
\\
{\bf Mass Enhancement.}  Due to the multi-band nature of  electronic structure in $\delta$-Pu, the cyclotron mass is not an effective comparison of mass enhancement due to the overall correlation induced renormalization. While the total number of electrons is conserved within each calculation method, the electron- and hole-like Fermi pockets expand when correlation effects are introduced, so in order to perform a meaningful comparison of the effective masses, a thermodynamic analysis of correlation-induced self-energies of Pu-5$f$ electrons is required~\cite{1997Kanki}. This is beyond the GGA and GGA+{\it U} methods, which are limited to the single-particle Kohn-Sham formalism.  We have done the comparison for the GGA+GutzA and GGA+DMFT methods using
\begin{equation}
\frac{m^{*}}{m_{b}} = \frac{\tilde{\rho} (E_{F})}{\rho_{b} (E_{F})},
\label{mass_enhanc}
\end{equation}
where $\rho_{b} (E_{F})=\sum_{j}w_{j}\rho_{b,j}(E_{F})$ is the band DOS and $\tilde{\rho} (E_{F})=\sum_{j}w_{j}\rho_{b,j}(E_{F})/z_{j}$ is the quasiparticle DOS. The partial density of states, $\rho_{b,j}(E_{F})$, is from the 14 5{\it f}-electron  spin orbitals, and indices $b$ and $j$ are the band index and spin quantum number respectively. In the case of the 5{\it f}-electrons, $j$ is either 5/2 or 7/2 where $w_{j}$ is the number of electrons with either spin (6 of $j=5/2$ and 8 of $j=7/2$).  In the GGA+DMFT method, the quasiparticle weight is $z_{j}=[1-\partial \textrm{Im}\Sigma_{j}(\omega_{n})/\left. \partial \omega_{n}\right\rvert_{\omega_{n}\rightarrow0}]^{-1}$, where $ \textrm{Im}\Sigma_{j}(\omega_{n})$ is the imaginary part of the electronic self-energy in terms of the Matsubara frequency, $\omega_{n}$~\cite{2014Zhu, 2016Zhu}.  In the GGA+GutzA method, the quasiparticle weights are the elements, $z_{j}$, of the matrix $Z_{j}\equiv R^{\dagger}_{j}R_{j}$, where the matrix $R_{j}$ is defined from the formulation of the rationally invariant slave boson theory (RISB) derived by Lanat\`a {\it et al.}\cite{2017Lanata}. We estimate the effective mass for the GGA+GutzA calculation to be $m^{*}= 1.30m_{b},$ and $m^{*}= 5.04m_{b}$ for the GGA+DMFT calculation. This is consistent with correlation induced band renormalization theory~\cite{1993Zwicknagl, 2016Zwicknagl}, which finds that effective mass is enhanced, overall, by strongly correlated electron-electron interactions.\\
\\
{\bf Temperature effects.} We have also explored the temperature dependence of the Fermi surface topology and electronic structure by using the GGA+DMFT method. Table~\ref{dHvATdep} contains the dHvA results obtained from electronic structure data for three temperatures.  
 As the temperature increased from 0 K to 116 K the extremal frequency of the hole bands (\#15,16) decreased and that of the electron bands increased. From 116 K to 1160 K the band structure evolves significantly such that only two degenerate bands (\#17,18) intersect the Fermi energy. The extremal frequency from this electron band isosurface is dramatically increased as is its reciprocal volume contained by the Fermi surface. It is noteworthy that the temperature T=1160 K is well above the melting point for Pu. These results are intended to demonstrate the range of the GGA+DMFT capabilities as well as explore the band renormalization in the region where Pu 5$f$-electrons become localized. Details of the electronic structure and Fermi surface topology at $T=1160$ K can be found in the SI document (Fig. S-8).
\begin{table*}
\caption{{\bf The dHvA and volume data for Fermi surface calculations at temperatures up to 1160 K.} Frequencies in kilotesla (kT), effective masses in units of $m_{e}$, and reciprocal Fermi surface volumes in units of $\AA^{-3}$ are shown for GGA+DMFT calculations of  $\delta$-Pu performed at $T=116$ K and $T=1160$ K . These are compared to the extrapolated DMFT calculations for $T = 0$ K.\label{dHvATdep}}
\begin{center}
\begin{tabular}{| l | r | r | r | r | r | r | r | r | r | r | r | r |}
\hline
&\multicolumn{3}{c|}{T=0 K} &\multicolumn{3}{c|}{T=116 K} &\multicolumn{3}{c|}{T=1160 K} \\
\hline
Band & $f$ & $m^{*}$ & $V_{FS}$ & $f$ & $m^{*}$ & $V_{FS}$ & $f$ & $m^{*}$ & $V_{FS}$ \\ 
\hline
15, 16 & $3.07$ & $1.84$ & $0.95$ & $2.38$ & $1.56$ & $0.57$ & $-$ & $-$ & $-$ \\ 
\hline
17, 18 & $6.18$ & $2.17$ & $0.95$ & $7.02$ & $1.98$ & $1.14$ & $13.80$ & $2.36$ & $3.04$ \\
\hline
\end{tabular}
\end{center}
\end{table*}

\noindent
{\bf Conclusion}
\\
In this paper we provide for the first time detailed calculations of the Fermi surface and associated mass renormalizations of $\delta$-Pu. By using a comparison between the results of four different theoretical methods, we have found a significant impact of strong electronic correlations on the $\delta$-Pu Fermi surface. For example, using a conventional GGA band-structure method as our starting point, three other methods that include electronic correlation effects beyond GGA showed a Fermi surface volume increase between ~200-230\% on individual Fermi pockets, depending on the method. The correlated-electron formalisms, GGA+GutzA and GGA+DMFT, which take quantum entanglements in the electronic structure into account, were in a relatively good agreement with each other, with both providing a renormalized electronic band structure and enhanced effective masses. Given the similarity in the results between the two methods and the computational efficiencies present in the GGA+GutzA method, this suggests that the GGA+GutzA may be highly beneficial for future studies of Fermi surface properties in other more complex Pu allotropes and alloys that may be beyond the current computational capabilities of GGA+DMFT.\\ 
\indent We also found that each theoretical method has a unique and identifiable impact on the electronic structure of $\delta$-Pu. These differences in the nonmagnetic and paramagnetic Fermi surface topologies, along with the Fermi surface data calculated for the ferromagnetic and antiferromagnetic cases (see SI document), should provide useful theoretical input to help analyze  future  magnetic quantum oscillation measurements. Temperature effects, for which there have been only a few extremely limited studies for Pu so far~\cite{2003Dai, 2017Dorado}, are another interesting factor to influence the Fermi surface topologies. Our comparison of calculations at 116 K and 1160 K show a dramatic evolution of the electronic structure and, consequently, the Fermi surface. Further studies on the temperature dependence of electronic correlation effects will be beneficial in advancing our understanding of Pu and other actinides.

\noindent
{\bf Methods}
\\
\\
{The $\delta$-Pu Fermi surface topologies were obtained using four electronic structure methods for comparison.}
\\
\\
\small
{\bf Density functional theory calculations.} Our starting point is conventional density functional theory (DFT) in the generalized gradient approximation (GGA) as implemented in the full-potential linearized augmented plane wave (FP-LAPW) method of WIEN2k~\cite{wien2k}. Relativistic spin-orbit coupling (SOC) effects were included by using a $k$-point grid of $15\time15\times15$ and a muffin-tin radius of 2.50$a_{0}$, where $a_{0}$ is the Bohr radius. These parameters were used for the basis of each calculation method.\\
\\
{\bf GGA+{\it U} calculations.}  GGA+{\it U} calculations were performed for non-magnetic $\delta$-Pu using the same parameters and exchange correlation functional as those used in the DFT method. The Hubbard parameter for the on-site Coulomb interaction strength was tested for values up to $U=4.5$ eV with an exchange-site parameter of $J=0.512$ eV, which has been shown to reproduce characteristics of $\delta$-Pu consistent with atomic spectral data~\cite{2006Kotliar}. The full electronic structure study for $U=0$ eV to $U=4.5$ eV (with $J=0.512$ eV for $U>0$) is available in the SI document. \\
\indent In addition to the non-magnetic cases, ferromagnetic and antiferromagnetic long-range magnetic ordering cases were explored using the GGA and GGA+{\it U} methods. The results can be found in the SI document. These cases were included purely for comparison as there has been no experimental evidence of long-range magnetic ordering detected in any phase of Pu.\\
\\
{\bf Gutzwiller approximation calculations.} Similar calculations were performed for paramagnetic $\delta$-Pu using the Gutzwiller approximation method implemented in the CyGutz code~\cite{2015Lanata, 2017Lanata}. This method is built upon an FP-LAPW based DFT~\cite{wien2k} calculation  and implements a combination of the slave-boson Gutzwiller wavefunction  method (GutzA) to account for strong electronic correlation. The Coulomb interaction strength was tested at values up to $U=4.5$ eV to facilitate comparison to the GGA+{\it U} and GGA+DMFT (details below) calculations. The electronic structures for $U=0$ eV to $U=4.5$ eV (with $J=0.512$ eV for $U>0$) is also included in the SI document. \\
\\
{\bf Dynamical mean-field theory calculations.} GGA+DMFT calculations were also performed for paramagnetic $\delta$-Pu. As in the case of the GGA+{\it U} and GGA+GutzA methods, this method uses the FP-LAPW implementation of WIEN2k~\cite{wien2k} as its basis. The DMFT calculation implements a strong-coupling version of the continuous-time quantum Monte-Carlo (CT-QMC) method~\cite{2007Zhu, 2010Haule, 2014Zhu} in order to explicitly consider the on-site Coulomb interactions with strength $U=4.5$ eV and $J=0.512$ eV. The remaining Slater integrals $F^2 = 6.1$ eV, $F^4 = 4.1$ eV and $F^6 = 3.0$ eV were calculated using Cowan’s atomic-structure code \cite{1981Cowan} and reduced by 30\% to account for screening.
 Calculations were performed at temperatures of $T=116$ K and $T=1160$ K. In order to compare the DMFT results to those of the other methods, for which $T=0$ K, the Fermi surface data was extrapolated from the $T=0.01$ eV calculation, where the sensitivity of the electronic self-energy  is negligible. The Fermi energy was then shifted to maintain the conservation of total number of electrons consistent with $T\rightarrow0$ eV. \\
\\
{\bf De Haas van Alphen calculations.} Analysis of the Fermi surfaces from each of the above methods was done using numerical calculations of the dHvA effect as implemented by Rourke and Julian~\cite{2012Rourke}. By applying a magnetic field to the system, oscillations in the magnetic susceptibility can be determined from the changes in the number of occupied Landau levels as a function of the reciprocal magnetic field, $1/\bf{B}$~\cite{2012Rourke, 2015Jiao}. Then the dHvA frequency can be expressed as\\
\begin{equation}
f_{i} = \frac{1}{\Delta (1/B)} = \frac{\hbar}{2\pi e} A_{i}
\label{dHvA_freq}
\end{equation}
where $e$ is the elementary charge of an electron, and $A_{i}$ is the extremal cross-sectional area of the $i^{th}$ branch of the Fermi surface in a plane perpendicular to $\bm{B}$. The effective carrier mass averaged around the extremal cyclotron orbits is also determined from
\begin{equation}
m^{*} = \frac{\hbar^{2}}{2\pi e}\left. \frac{\partial A}{\partial E}\right\rvert_{E=E_{F}}
\label{dHvA_mass}
\end{equation}
where $m^{*}$ is in units of the electron mass, $m_{e}$.\\ 
\indent The results of the dHvA analysis (Table~\ref{dHvAT0}) are to be compared against magnetic quantum oscillation measurements. We anticipate that such measurements will aid in determining the correct model to describe the physical properties of Pu.\\

\vspace{0.5cm}

\noindent{\bf Acknowledgements~}
We thank Filip Ronning, Qimiao Si, John Singleton, Peter Wolfle, N. Harrison, J. M. Wills, and Gertrud Zwicknagl for helpful discussions. This work was carried out under the auspices of the U.S. Department of Energy (DOE) National Nuclear Security Administration under Contract No. 89233218CNA000001. The Fermi surface topology analysis work was supported by LANL LDRD  Program. The DFT+DMFT simulation work at high temperature was supported by the NNSA Advanced Simulation and Computing Program. It was in part supported by Center for Integrated Nanotechnologies, a DOE BES user facility, in partnership with LANL Institutional Computing Program for computational resource. 

\vspace{0.5cm}
\noindent{\bf Correspondence} and requests for materials should be
addressed to R.M.T. (rtutchton@lanl.gov) or J.-X.Z. (jxzhu@lanl.gov).

%\bibliographystyle{apsrev}
%\bibliography{FS_top}

\end{document}

\title{Supplementary Information: Electronic Correlation Induced Expansion of Compensated Electron and Hole Fermi Pockets in $\delta$-Plutonium}

\author{Roxanne Tutchton}
\email{rtutchton@lanl.gov}
\affiliation{Theoretical Division, Los Alamos National Laboratory, Los Alamos, New Mexico 87545, USA}

\author{Wei-ting Chiu}
\affiliation{Department of Physics, University of California,  Davis, California 95616, USA}

\author{R. C. Albers}
\affiliation{Theoretical Division, Los Alamos National Laboratory, Los Alamos, New Mexico 87545, USA}

\author{G. Kotliar} 
\affiliation{Department of Physics and Astronomy, Rutgers University, Piscataway, New Jersey 08854, USA}

\author{Jian-Xin Zhu}
\email{jxzhu@lanl.gov}
\affiliation{Theoretical Division, Los Alamos National Laboratory,
Los Alamos, New Mexico 87545, USA}
\affiliation{Center for Integrated Nanotechnologies, Los Alamos National Laboratory,
Los Alamos, New Mexico 87545, USA}

%\pacs{71.27.+a, 74.55.+v, 75.20.Hr, 75.30.Mb}
%71.27.+a	Strongly correlated electron systems; heavy fermions
%74.55.+v	Tunneling phenomena: single particle tunneling and STM
%75.20.Hr	Local moment in compounds and alloys; Kondo effect, valence fluctuations, heavy fermions
%75.30.Mb	Valence fluctuation, Kondo lattice, and heavy-fermion phenomena
\maketitle

%\onecolumngrid
%\twocolumngrid
%\section*{Supplemental Material for EPAPS}

%\begin{equation}
%x=y
%\end{equation}
%
%\begin{subequations}
%\begin{eqnarray}
%A&=&B\;, \\
%C&=& D\;.
%\end{eqnarray}
%\end{subequations}

%\begin{figure}[h!]
%\centering
%\subfigure[\ Real and imaginary.]{
%\includegraphics[scale=0.5]{reim}}
%\subfigure[\ Amplitude and phase.]{
%\includegraphics[scale=0.5]{phase}}
%\caption{\label{fig:qm/complexfunctions} Two representations of complex
%wave functions.}
%\end{figure}

%\onecolumngrid
%\twocolumngrid

In the following Supplemental Information (SI), we show details for and additions to several of the calculations presented in the main text. We include GGA and GGA+$U$ calculations with long-range magnetic ordering in $\delta$-Pu as well as on-site Coulomb parameter analysis for the GGA+$U$ and GGA+GutzA methods. Details of our temperature dependent study of $\delta$-Pu using DMFT are included, and an analysis of the ${\bf B}$-field angular dependence in the dHvA simulations is given for both zero temperature calculations and the temperature dependent DMFT calculations. The additional information presented in this SI document gives the broader context of our calculations and the extra details we used to arrive at the conclusions in the main text.

\section{Long-range magnetic ordering electronic structure calculations}

\begin{figure}[th]
\centering\includegraphics[%scale=0.28
width=1.0\linewidth,clip]{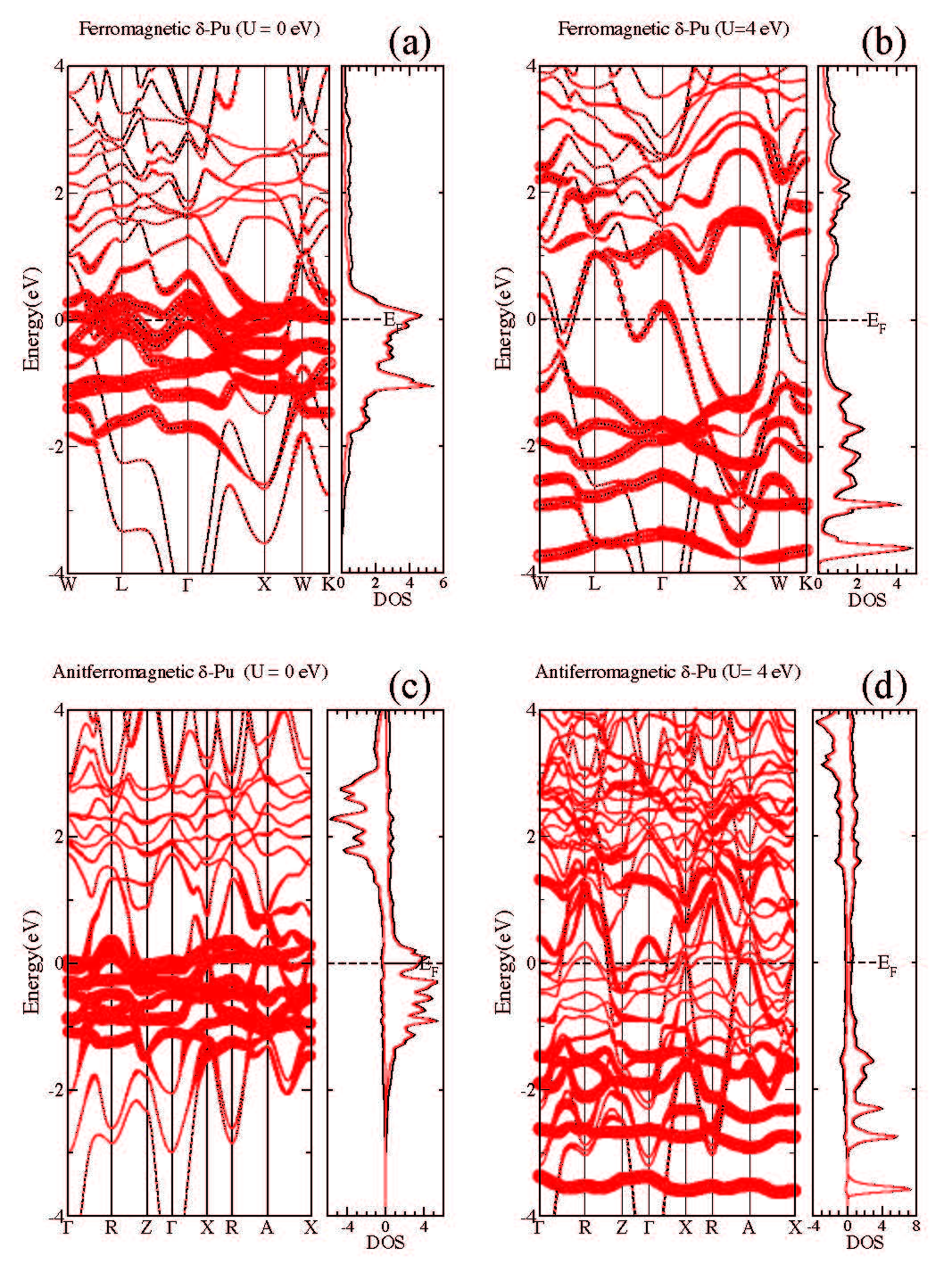}
\caption{(Color online) 
The electronic band structures and DOS for $\delta$-Pu with (a) ferromagnetic ordering for $U=0$ eV, (b) ferromagnetic ordering for $U=4.5$ eV and $J =0.512$ eV, (c) A-type antiferromagnetic ordering for $U=0$ eV, and (b) A-type antiferromagnetic ordering for $U=4.5$ eV and $J =0.512$ eV. The thick bands of the electronic dispersions indicate $f$-electron occupation. Likewise, the red curve of the density of states indicates the f-projected DOS, and the black curves show the total DOS in units of states/eV/unit cell. For the DOS in the antiferromagnetic cases, (c) and (d), the DOS is given for both spin-up (positive) and spin-down (negative) states.
}
\label{Sfig:BS_MagOrd}
\end{figure}

\begin{figure}[th]
\centering\includegraphics[%scale=0.28
width=1.0\linewidth,clip]{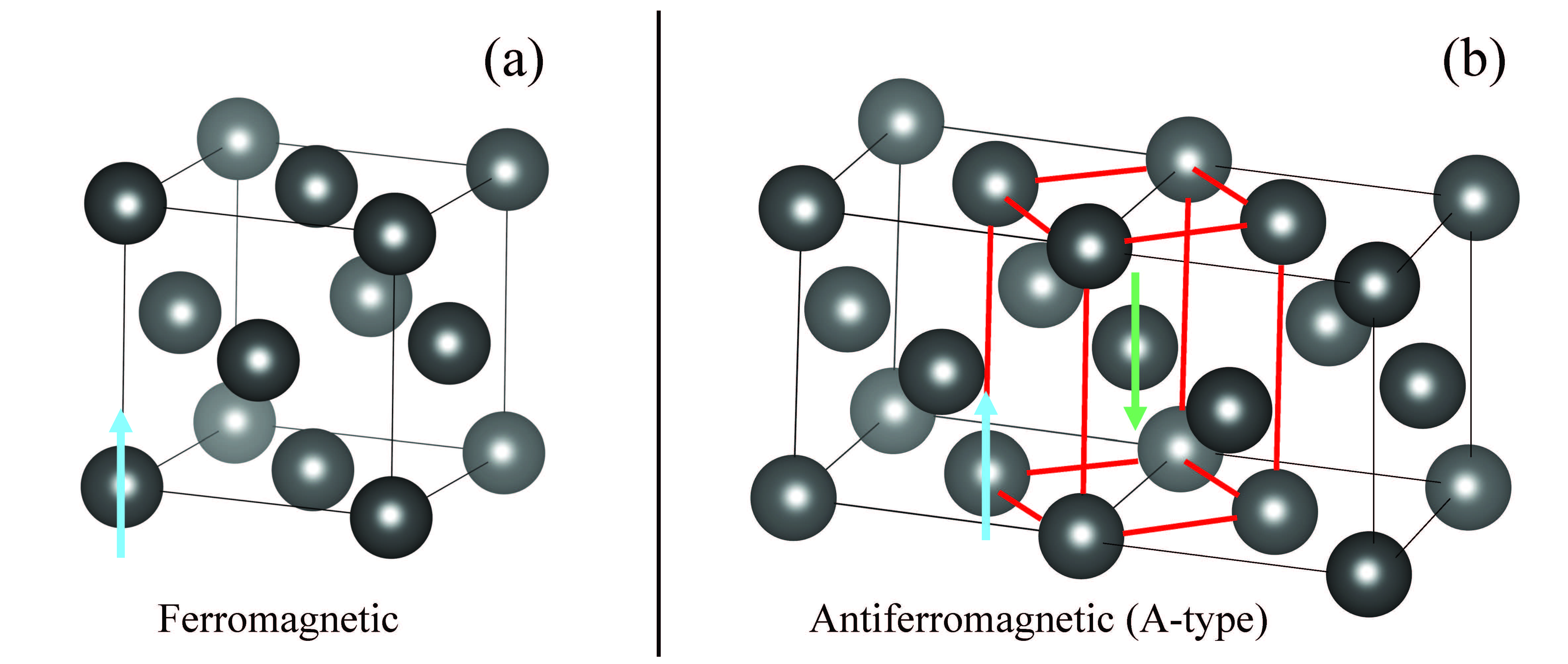}
\caption{(Color online) 
The long-range magnetic ordering configurations assumed for $\delta$-Pu. (a) The ferromagnetic ordering is described in the standard fcc structure for $\delta$-Pu. The blue arrow indicates the spin-up configuration. As there is only one atom in the primitive fcc cell, all of the atoms are defined to have a spin-up orientation. (b) For the antiferromagnetic case, the primitive cell is reconfigured into a body-centered-tetragonal (bct) symmetry, which has two atoms in its primitive cell. These are specified to be spin-up (blue) and spin-down (green) giving an A-type antiferromagnetic configuration. 
}
\label{Sfig:MagOrd_FandAtype}
\end{figure}

\begin{figure*}[th]
\centering\includegraphics[%scale=0.28
width=0.8\linewidth,clip]{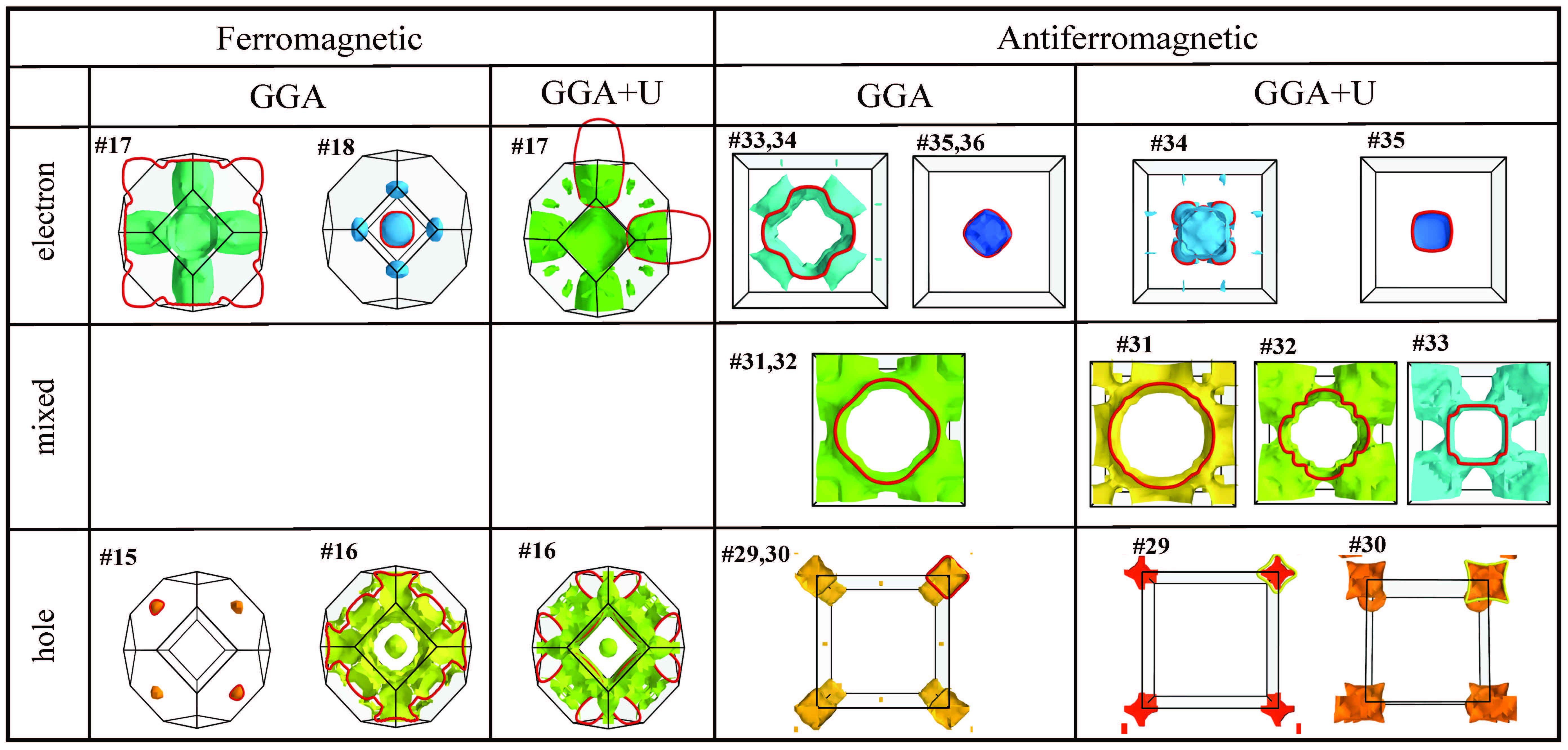}
\caption{(Color online) 
The Fermi surfaces for assumed ferromagnetic and antiferromagnetic $\delta$-Pu. The isosurfaces are arranged according to the calculation method – GGA and GGA+$U$ (for $U = 4.5$ eV and $J=0.512$ eV) and carrier type—electron, hole, and mixed (both electrons and hole quasiparticles). The band number is indicated near each surface and the red curves (yellow for bands 29 and 30 of the antiferromagnetic GGA+$U$ calculation) indicate the cross-sectional area associated with the extremal dHvA frequency listed in Table I and Table II.  The Fermi surfaces are oriented such that the cartesian z axis is perpendicular to the outlined cross-sectional areas (see Fig. 3b of the main text). The applied magnetic field of the dHvA calculations is parallel to the $z$-axis (${\bf B}||z$). 
}
\label{Sfig:FS_MagOrd}
\end{figure*}

Since there are still some concerns about the question of the magnetic ordering present in Pu\cite{2006Heffner, 2005Lashley, 2017Janoschek, 2017Migliori}, we provide, for comparison, calculated electronic structures, Fermi surface topologies, and de Haas-van Alphen (dHvA) results for the assumed ferromagnetic and antiferromagnetic $\delta$-Pu. The calculations were performed by using both standard DFT in the generalized gradient approximation (GGA) and GGA+$U$, with the Coulomb strength parameter $U = 4.5$ eV and exchange parameter $J=0.512$ eV.  Figure~\ref{Sfig:BS_MagOrd} shows the electronic band structures and density of states (DOS) for each case. The ferromagnetic ordering is calculated using the standard $\delta$-Pu face-center-cubic (fcc) symmetry; while a body-centered-tetragonal configuration, containing two atoms in its unit cell, is used for the antiferromagnetic case, to account for the oppositely aligned spins, as shown in Fig.~\ref{Sfig:MagOrd_FandAtype}.\\
 \indent The corresponding ferromagnetic and antiferromagnetic Fermi surfaces are given in Fig.~\ref{Sfig:FS_MagOrd}, and the dHvA frequencies, cyclotron effective masses, and reciprocal volumes are listed in Tables I and II for the GGA and GGA+$U$ methods, respectively. It is notable that the Fermi surface topologies and dHvA data for the non-magnetic/paramagnetic case and those with long-range magnetic ordering are vastly different from one another. The addition of the $U$ parameter in the magnetic cases causes splitting of the degenerate energy bands as well as significant shifting. This is evident in both the band dispersions and the Fermi surface topologies. Figure~\ref{Sfig:BS_MagOrd} panels (a) and (b) reveal a dramatic evolution in the electronic dispersion of ferromagnetic $\delta$-Pu when the Coulomb interactions are included. The same observation can be made for panels (c) and (d) showing the antiferromagnetic case. Furthermore, these electronic structures are distinct from those found for the non-magnetic/paramagnetic cases discussed in the main text. These additional results will be useful for the comparisons to magnetic quantum oscillation measurements when they are available, perhaps helping to clear up some of the questions concerning possible long-range order in $\delta$-Pu.

\begin{table}
\label{T:GGA_dHvA}
\caption{The dHvA data for the GGA method are given for each band in the ferromagnetic and antiferromagnetic cases. The extremal frequencies are in units of kilotesla (kT), the cyclotron masses are in terms of the electron mass ($m_{e}$), and the reciprocal volume is given in inverse cubic angstroms ($\AA^{-3}$).}
\begin{center}
\begin{tabular}{| l | c | c | c | c | c | c |}
\hline
& \multicolumn{3}{c|}{Ferromagnetic} & \multicolumn{3}{c|}{Antiferromagnetic}\\
\hline
Band & $f$(kT) & $m^{*}(m_{e})$ & $V_{FS}(\AA^{-3})$ & $f$(kT) & $m^{*}(m_{e})$ & $V_{FS}(\AA^{-3})$\\ 
\hline
15 & $0.56$ & $1.17$ & $0.03$ &-&-&-\\ 
16 & $36.6$ & $30.2$ & $3.22$ &-&-&- \\ 
17 & $58.4$ & $23.2$ & $2.87$ &-&-&- \\
18 & $3.05$ & $2.15$ & $0.38$ &-&-&- \\
\hline
29 & - & - & - & $2.20$ & $1.57$ & $0.06$\\ 
30 & - & - & - & $2.20$ & $1.57$ & $0.06$\\ 
31 & - & - & - & $12.9$ & $4.47$ & $1.40$\\
32 & - & - & - & $12.9$ & $4.47$ & $1.40$\\
33 & - & - & - & $8.91$ & $4.67$ & $1.23$\\ 
34 & - & - & - & $8.91$ & $4.67$ & $1.23$\\ 
35 & - & - & - & $2.86$ & $4.18$ & $0.24$\\
36 & - & - & - & $2.86$ & $4.18$ & $0.24$\\
\hline
\end{tabular}
\end{center}
\end{table}

\begin{table}
\label{T:GGA+U_dHvA}
\caption{The dHvA data for the GGA+$U$ method are given for each band in the ferromagnetic and antiferromagnetic cases. The extremal frequencies are in units of kilotesla (kT), the cyclotron masses are in terms of the electron mass ($m_{e}$), and the reciprocal volume is given in inverse cubic angstroms ($\AA^{-3}$).}
\begin{center}
\begin{tabular}{| l | c | c | c | c | c | c |}
\hline
& \multicolumn{3}{c|}{Ferromagnetic} & \multicolumn{3}{c|}{Antiferromagnetic}\\
\hline
Band & $f$ & $m^{*} $ & $V_{FS}$ & $f$ & $m^{*}$ & $V_{FS}$\\ 
\hline
16 & $25.9$ & $6.94$ & $2.79$ &-&-&- \\ 
17 & $13.8$ & $1.31$ & $2.79$ &-&-&- \\
\hline
29 & - & - & - & $1.55$ & $2.27$ & $0.01$\\ 
30 & - & - & - & $3.71$ & $3.15$ & $0.05$\\ 
31 & - & - & - & $15.4$ & $5.51$ & $0.73$\\
32 & - & - & - & $10.5$ & $5.05$ & $1.92$\\
33 & - & - & - & $5.81$ & $1.14$ & $2.95$\\ 
34 & - & - & - & $6.97$ & $11.9$ & $0.56$\\ 
35 & - & - & - & $2.45$ & $1.63$ & $0.12$\\
\hline
\end{tabular}
\end{center}
\end{table}

\section{Coulomb interaction study in $\delta$-Pu. }
\begin{figure*}[t]
\centering\includegraphics[%scale=0.28
width=0.8\linewidth,clip]{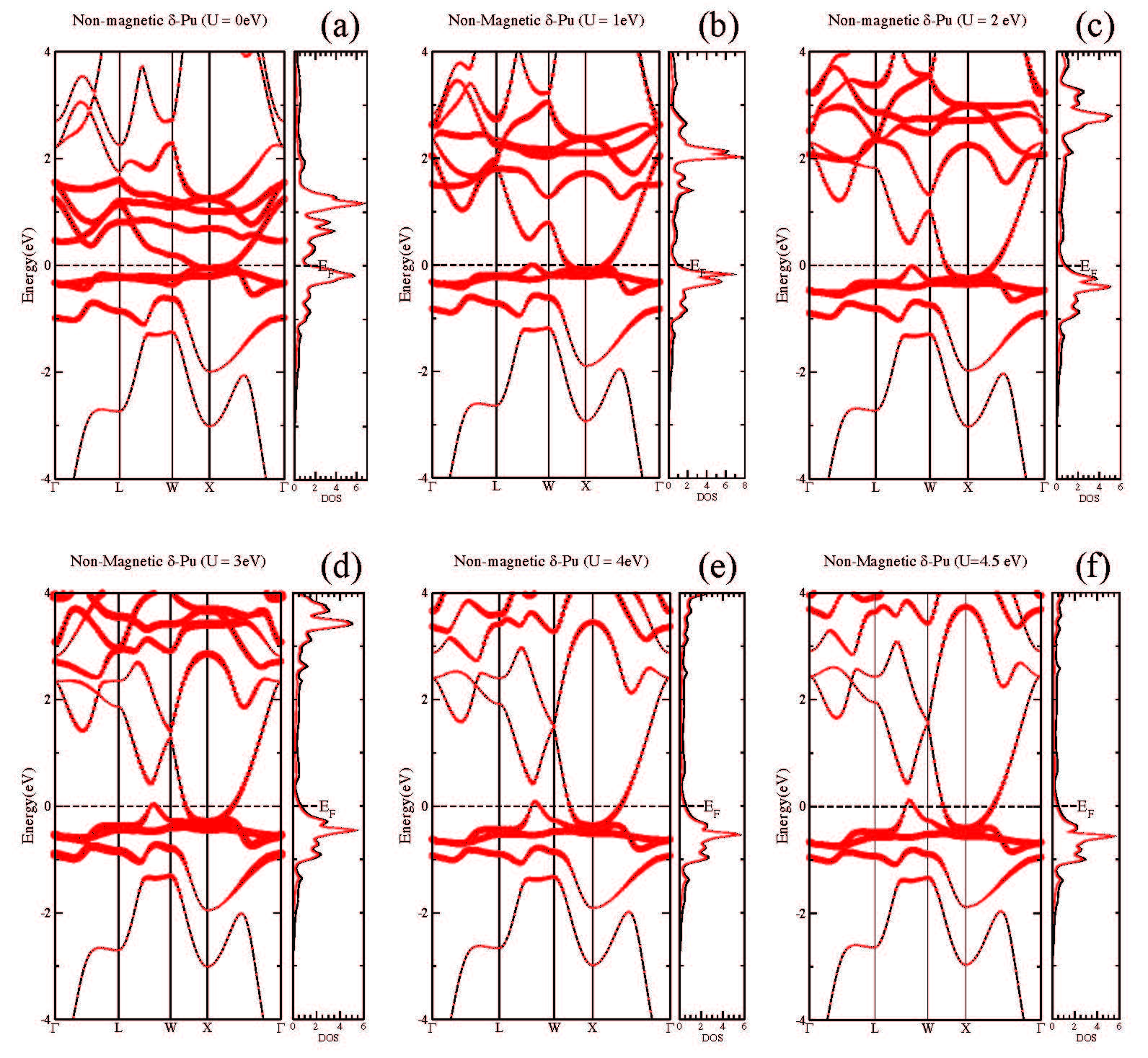}
\caption{(Color online) 
The electronic band structures and DOS for $\delta$-Pu calculated in the GGA+$U$ for $U= 0$ to 4.5 eV. Panels (a) through (f) show the band dispersion and DOS in states/eV for increasing values of $U$. The thick red bands of the electronic dispersions and red curve on the DOS indicate the $f$-electron occupation and the $f$-projected DOS, respectively. The black curves on the DOS show the total density.
}
\label{Sfig:GGA+U_study}
\end{figure*}

\begin{figure*}[tbh]
\centering\includegraphics[%scale=0.28
width=0.8\linewidth,clip]{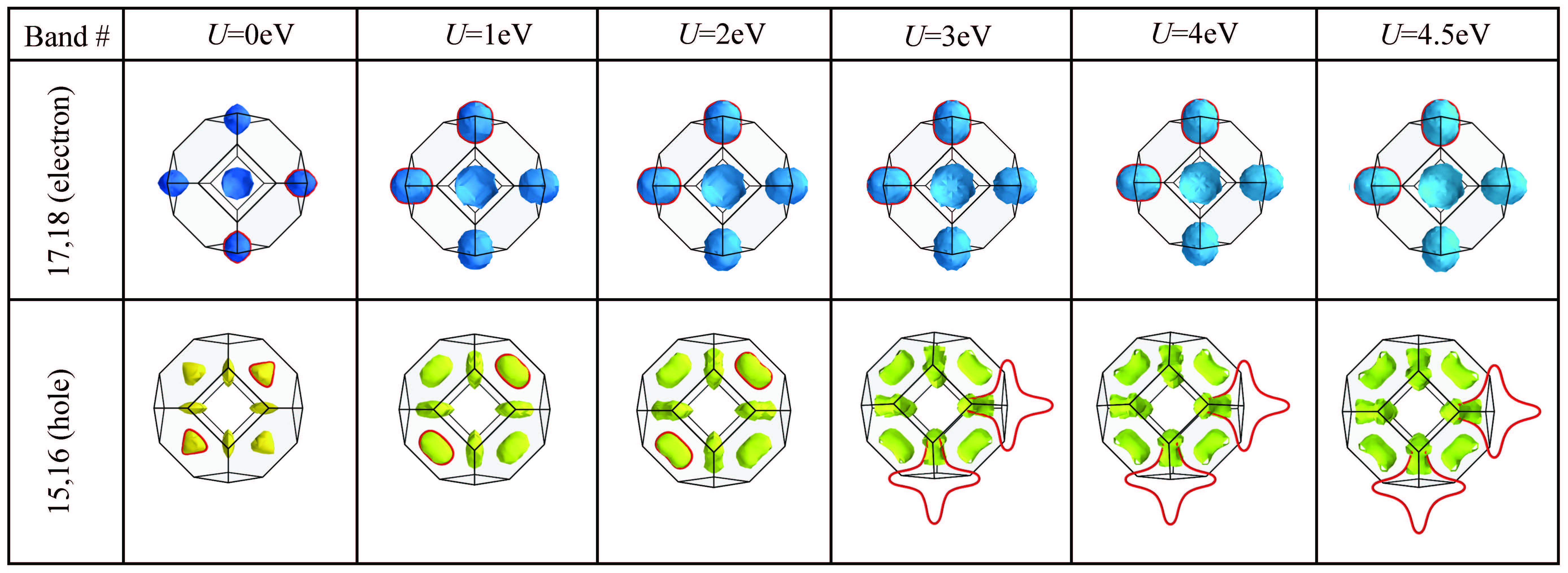}
\caption{(Color online) 
A study of increasing $U$ in the Fermi surfaces for non-magnetic $\delta$-Pu calculated with the GGA+$U$ method. The isosurfaces are arranged by carrier type. The red curves indicate the cross-sectional area for ${\bf B}||z$ associated with the extremal dHvA frequency listed in STable III. Each Fermi surfaces is oriented such that the $z$-axis is perpendicular to the outlined cross-sectional area (see Fig. 3b of the main text). For $U\geq 3$ the extremal area outlined (in red) by the dHvA frequency extends past the 1st Brillouin zone resulting in a large increase in the extremal frequency (Table III).
}
\label{Sfig:FS_GGA+Ustudy}
\end{figure*}

The Coulomb strength was tested for several values of $U$ using both GGA+$U$ calculations and GGA+GutzA calculations. The evolution of the electronic band structure and DOS for the GGA+$U$ calculations is shown in Fig.~\ref{Sfig:GGA+U_study}, the corresponding Fermi surfaces are given in Fig.~\ref{Sfig:FS_GGA+Ustudy}., and the dHvA data are listed in Table III. As $U$ is increased from 0 eV to 4.5 eV, the hole and electron pockets around the Fermi energy expand, and the bands shift resulting in a concentration of $f$-electrons from the $j=5/2$ subshell just below the fermi energy and $j=7/2$ subshell electrons between 4 and 6 eV. Similarly, the expansion of the electron and hole Fermi surfaces is apparent in Fig.~\ref{Sfig:FS_GGA+Ustudy} as well as the frequency and volume results in Table III. The exchange parameter $J = 0.512$ eV for all calculations where $U > 0$.

\begin{table}
\caption{The dHvA data in the GGA+$U$ method for $U=0$ to 4.5 eV and $J=0.512$ eV are given for each band. The extremal frequencies are in units of kilotesla (kT), the cyclotron masses are in terms of the electron mass ($m_{e}$), and the reciprocal volume is given in inverse cubic angstroms ($\AA^{-3}$). As there are two sets of degenerate bands crossing the Fermi energy, the reciprocal volumes show the expansion in electron pocket and the compensation in the hole pocket.}
\begin{center}
\begin{tabular}{| c | c | c | c | c | c | c |}
\hline
Band &  \multicolumn{3}{c|}{15,16} & \multicolumn{3}{c|}{17,18} \\ 
\hline
$U$ (eV) & $f$ & $m^{*}$ & $V_{FS}$ & $f$ & $m^{*}$ & $V_{FS}$\\ 
\hline
0 & $1.73$ & $3.19$ & $0.31$ & $2.86$ & $5.36$ & $0.31$ \\ 
1 & $3.01$ & $2.02$ & $0.78$ & $5.62$ & $2.81$ & $0.78$ \\
2 & $2.60$ & $1.72$ & $0.90$ & $4.69$ & $3.21$ & $0.90$ \\
3 & $10.1$ & $2.92$ & $0.96$ & $6.42$ & $1.84$ & $0.96$ \\
4 & $10.2$ & $2.35$ & $1.00$ & $6.66$& $1.54$ & $1.00$ \\
4.5 & $10.2$ & $2.35$ & $1.02$ & $6.66$ & $1.54$ & $1.02$ \\
\hline
\end{tabular}
\end{center}
\end{table}

\begin{figure*}[tbh]
\centering\includegraphics[%scale=0.28
width=0.8\linewidth,clip]{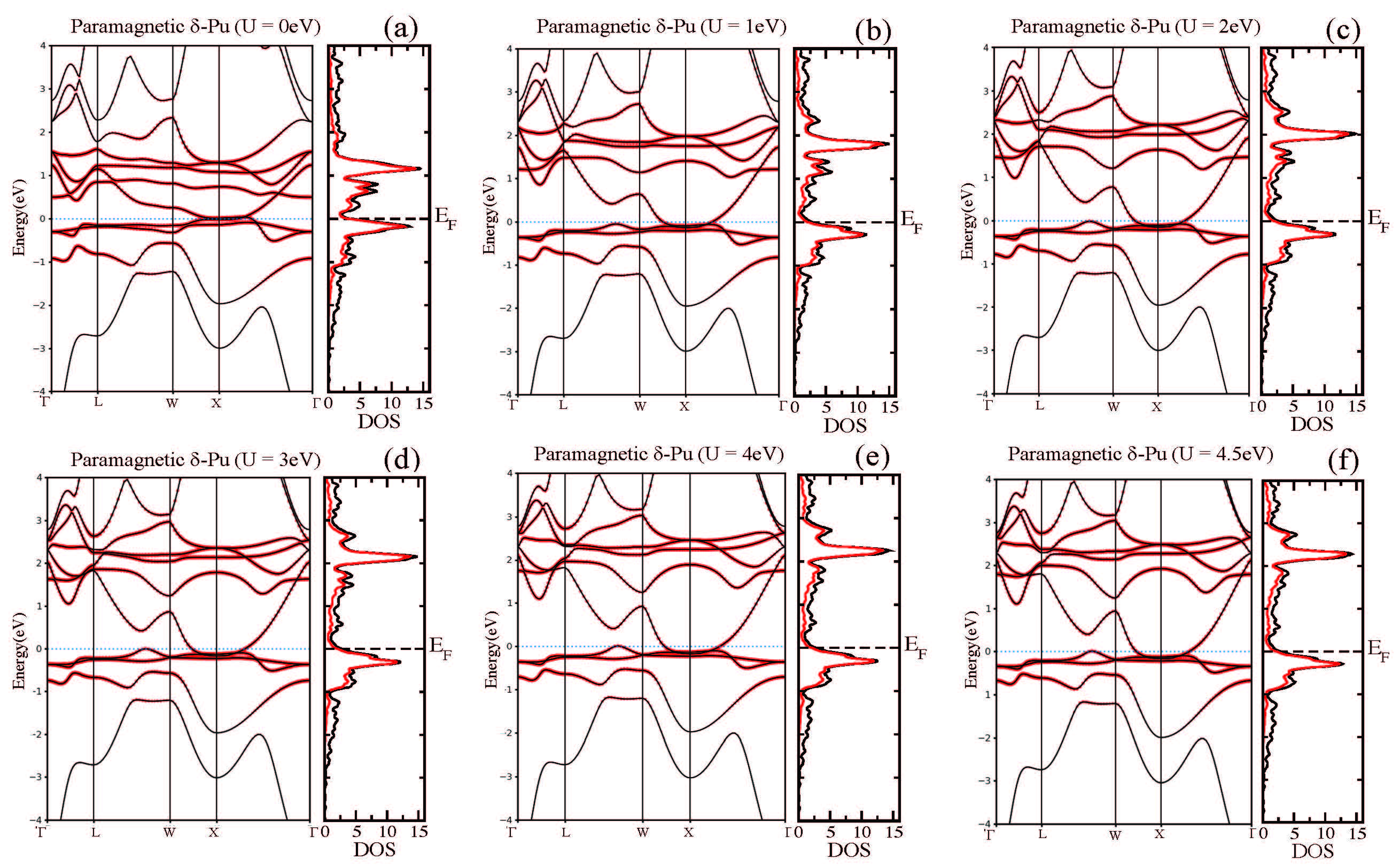}
\caption{(Color online) 
The electronic band structures and DOS for $\delta$-Pu calculated in the GGA+GutzA for $U= 0$ to 4.5 eV. Panels (a) through (f) show the band dispersion and DOS in states/eV for increasing values of $U$. The thick red bands of the electronic dispersions and red curve on the DOS indicate the f-electron occupation and the $f$-projected DOS, respectively. The black curves on the DOS show the total density.}
\label{Sfig:GGA+GutzAstudy}
\end{figure*}

\begin{figure*}[tbh]
\centering\includegraphics[%scale=0.28
width=0.8\linewidth,clip]{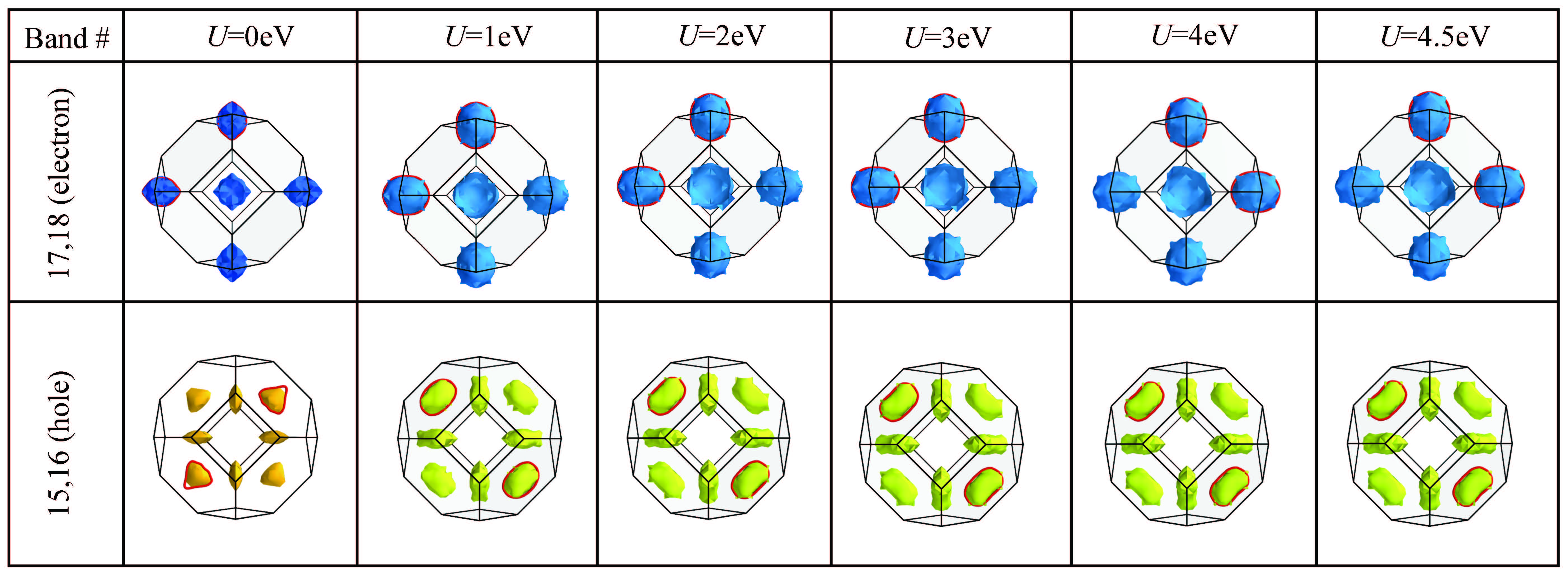}
\caption{(Color online) 
A study of the $U$-strength effect on the Fermi surfaces for non-magnetic $\delta$-Pu calculated with the GGA+GutzA method. The isosurfaces are arranged by carrier type. The red curves indicate the cross-sectional area for ${\bf B}||z$ associated with the extremal dHvA frequency listed in Table S-IV. Each Fermi surfaces is oriented such that the z axis is perpendicular to the outlined cross-sectional area (see Fig. 3b of the main text).}
\label{Sfig:FS_GGA+GutzAstudy}
\end{figure*}

The $U$-dependence analysis of the electronic structure presented in Fig.~\ref{Sfig:GGA+GutzAstudy} shows the same range of $U = 0$ to $U = 4.5$ eV for the GGA+GutzA calculations. Again, the exchange parameter $J = 0.512$ eV for all calculations where $U > 0$. There are several similarities between the GGA+$U$ and GGA+GutzA electronic structures. The development of the hole pocket between the high symmetry points $L$ and $W$ appears in both Figs. S4 and S6 as does the increasing separation between the $j=5/2$ and $j=7/2$ $f$-electron subshells. However, for the band structure calculated in the GGA+GutzA the shift in the conduction bands is significantly less pronounced. This is apparent in the Fermi surface development as well (Fig.~\ref{Sfig:FS_GGA+GutzAstudy}). The hole bands never extend past the first Brillouin zone (BZ), and there is, subsequently, no sharp increase in the dHvA extremal frequency for bands 15 and 16 in Table IV, as there is in Table III. 

\begin{table}
\caption{The dHvA data in the GGA+GutzA method for $U = 0$ to 4.5 eV are given for each band. The extremal frequencies are in units of kilotesla (kT), the cyclotron masses are in terms of the electron mass ($m_{e}$), and the reciprocal volume is given in inverse cubic angstroms ($\AA^{-3}$). As there are two sets of degenerate bands crossing the Fermi energy, the reciprocal volumes show the expansion in electron pocket and the compensation in the hole pocket.}
\begin{center}
\begin{tabular}{| c | c | c | c | c | c | c |}
\hline
Band &  \multicolumn{3}{c|}{15,16} & \multicolumn{3}{c|}{17,18} \\ 
\hline
$U$ (eV) & $f$ & $m^{*}$ & $V_{FS}$ & $f$(kT) & $m^{*}$ & $V_{FS}$\\ 
\hline
0 & $1.75$ & $3.79$ & $0.31$ & $2.88$ & $6.08$ & $0.31$ \\ 
1 & $2.95$ & $2.39$ & $0.76$ & $5.55$ & $2.93$ & $0.76$ \\
2 & $3.07$ & $2.16$ & $0.83$ & $5.92$ & $2.61$ & $0.83$ \\
3 & $3.11$ & $2.07$ & $0.88$ & $6.09$ & $2.53$ & $0.88$ \\
4 & $3.12$ & $2.01$ & $0.90$ & $6.22$& $2.48$ & $0.90$ \\
4.5 & $3.12$ & $2.01$ & $0.91$ & $6.27$ & $2.48$ & $0.91$ \\
\hline
\end{tabular}
\end{center}
\end{table}

\section{Temperature study in the DMFT method }
\begin{figure*}[tbh]
\centering\includegraphics[%scale=0.28
width=1.0\linewidth,clip]{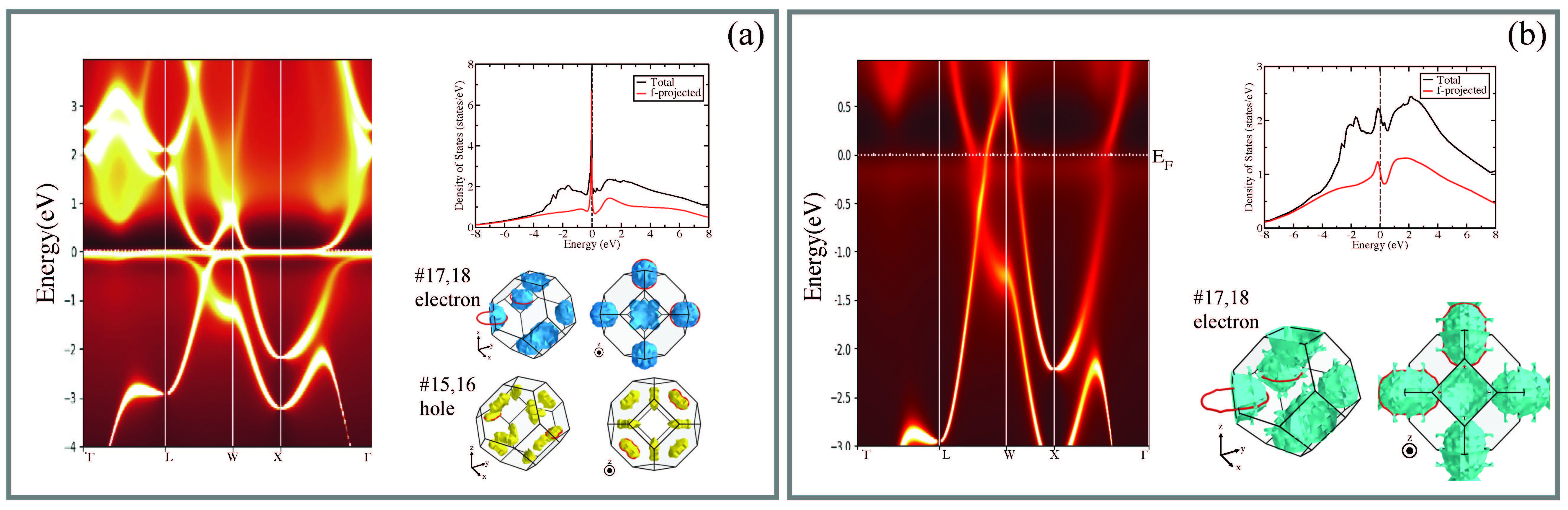}
\caption{(Color online) 
The electronic structures and Fermi surfaces for $\delta$-Pu at (a) $T=116$ K and (b) $T=1160$ K calculated using the GGA+DMFT method. Each panel shows the band dispersion, DOS in states/eV, and the Fermi surfaces corresponding to the band crossings. Notably, the high temperature calculation indicates that only the electron band crosses the Fermi energy, so there is only one Fermi surface. }
\label{Sfig:tempDMFT}
\end{figure*}

Figure~\ref{Sfig:tempDMFT} shows GGA+DMFT calculations for $T=116$ K and $T=1160$ K in order to explore the evolution of the Fermi surface topology at the extreme temperature limits of the calculation method. We note that such a high temperature (already above the melting temperature) is used to model the regime where Pu 5$f$-electrons are incoherent. The results indicate that the concentration of 5$f$-electrons near the Fermi energy at low temperatures is mostly dispersed to high energies at $T =1160$ K. The Fermi surface has also evolved to a single degenerate band crossing. The dHvA data are discussed in the main text.

\section{Angular analysis of dHvA frequencies and cyclotron effective masses }
\begin{figure*}[tbh]
\centering\includegraphics[%scale=0.28
width=0.7\linewidth,clip]{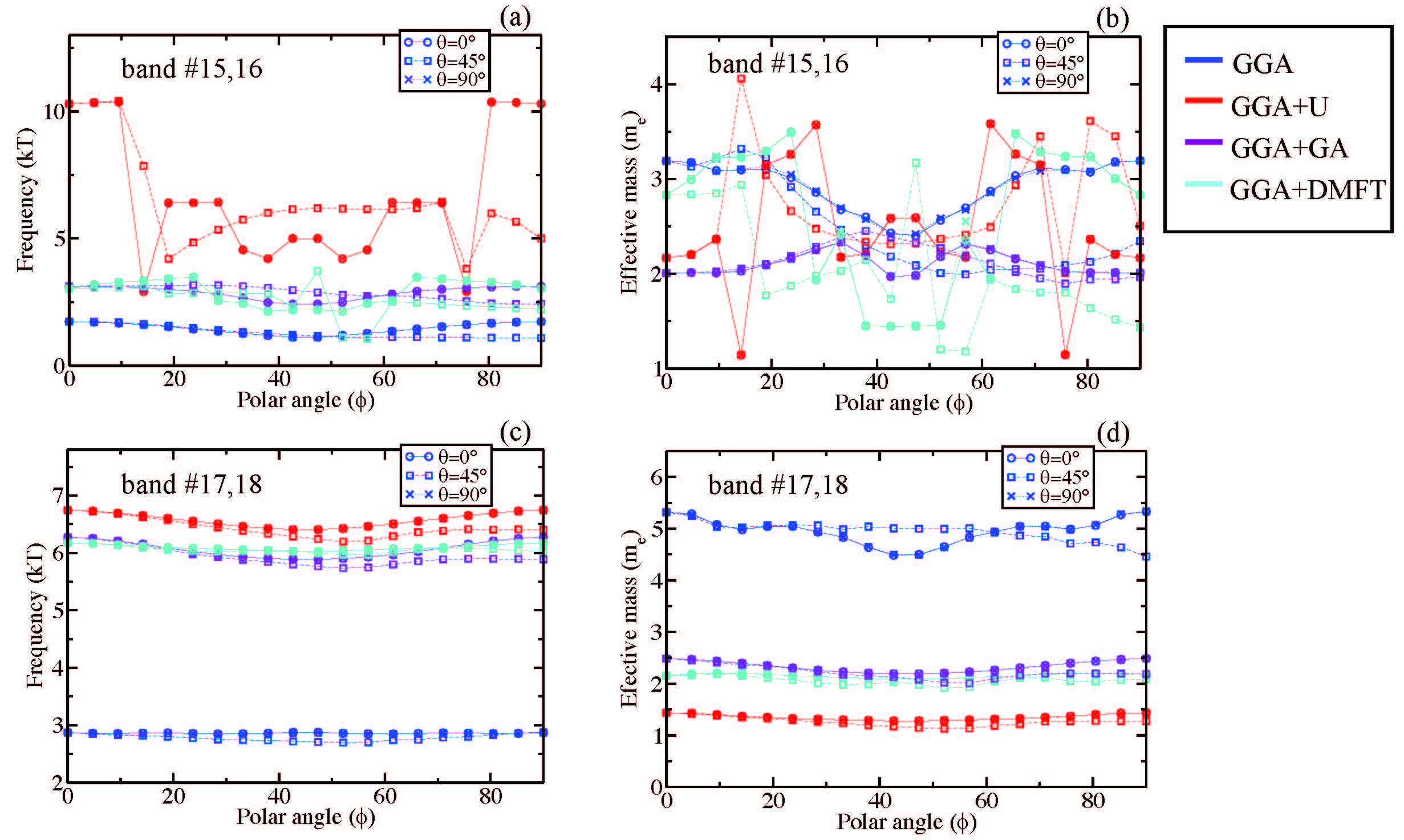}
\caption{(Color online) 
The ${\bf B}$-field angular dependence of the dHvA frequencies and cyclotron masses for $\delta$-Pu at $T=0$ K. Graphs (a) and (c) show the extremal frequency dependence on the ${\bf B}$-field angle (in polar coordinates) for degenerate band pairs 15\&16 and 17\&18, respectively. Graphs (b) and (d) show the same dependence for the cyclotron effective masses. The vertical axis on each graph gives the polar angle, $\phi$ (following the notation of Rourke and Julian~\cite{2012Rourke}). The symbols marking the data points change depending on the azimuthal angle, where the circles are for $\theta=0\degree$, the squares for $\theta=45\degree$, and the X’s for $\theta=90\degree$.  The colors correspond to the calculation methods where the GGA results are in blue, GGA+$U$ in red, GGA+GutzA in purple, and the DMFT in turquoise. The correlation methods (GGA+$U$, GGA+GutzA, and GGA+DMFT) are all done with $U = 4.5$ eV and $J=0.512$ eV. }
\label{Sfig:AngularT0}
\end{figure*}

Previously we have discussed extremal frequencies and cyclotron effective masses calculated from dHvA effects in the presence of an external magnetic field with one orientation, ${\bf B}||z$. Here we also show the results of a ${\bf B}$-angle dependence analysis for non-magnetic/paramagnetic $\delta$-Pu calculated using the four methods discussed in the main text. Figure~\ref{Sfig:AngularT0} shows the results of our $T=0$ K comparisons. The trends for most of the methods show minimal variation in the frequency results and only slight enhancement in the effective masses. The exception to this is the GGA+$U$ method. This can be seen most clearly in the hole band (Fig.~\ref{Sfig:AngularT0} (a) and (b)), which has large jumps in the frequency and mass curves for several azimuthal angles. An explanation for this can be seen in the Fermi surface for the GGA+$U$ hole band (Fig.~\ref{Sfig:FS_GGA+Ustudy}). Unlike the other methods, the \#15\&16 isosurface extends beyond the 1st BZ, resulting in a more complicated topology with larger variations in its dHvA data.  It is also evident, when comparing the calculation methods, that the GGA+GutzA and GGA+DMFT Fermi surfaces are in close agreement. 

\begin{figure*}[tbh]
\centering\includegraphics[%scale=0.28
width=0.7\linewidth,clip]{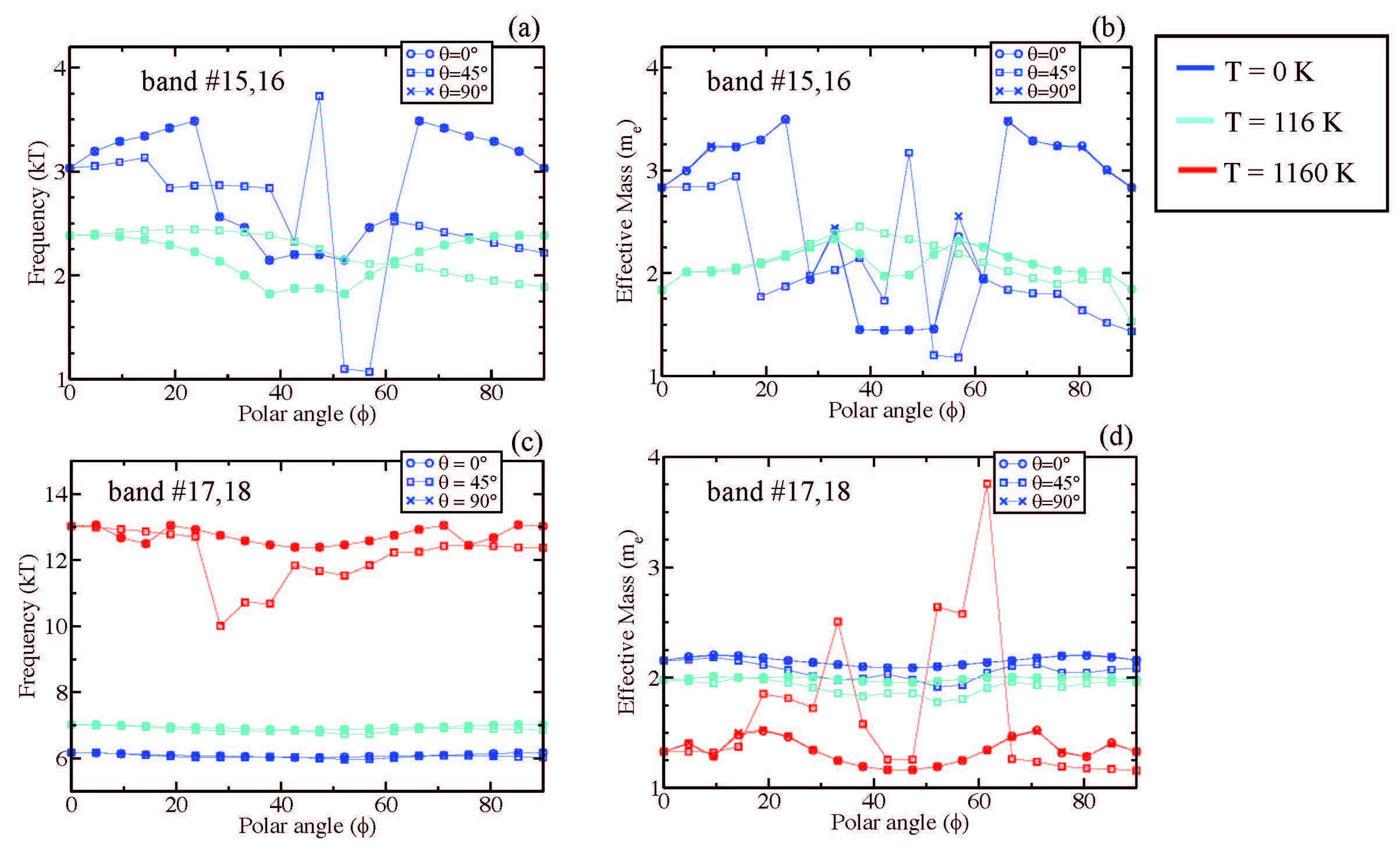}
\caption{(Color online) 
The temperature and B-field angular dependence of the dHvA frequencies and cyclotron masses for $\delta$-Pu from $T=0$ to 1160 K calculated using GGA+DMFT with $U = 4.5$ eV and $J=0.512$ eV. Graphs (a) and (c) show the extremal frequency dependence on the ${\bf B}$-field angle for degenerate band pairs 15\&16 and 17\&18, respectively. Graphs (b) and (d) show the cyclotron effective masses. The vertical axis on each graph gives the polar angle, $\delta$. The circles are for $\theta=0\degree$, the squares for $\theta=45\degree$, and the X’s for $\theta=90\degree$.  The colors correspond to the temperatures, where blue is $T=0$ K, turquoise is $T=116$ K, and red is $T=1160$ K.}
\label{Sfig:AngularTdeb}
\end{figure*}

Figure~\ref{Sfig:AngularTdeb} gives the angular comparison for the temperature study done with GGA+DMFT. Here we have compared the adjusted $T = 0$ K dHvA data to the DMFT results for 116 K and 1160 K. The change from 0 K to 116 K is relatively small, as is expected for the low temperature region. The increase from $T=116$ K to $T=1160$ K leads to a large frequency increase for bands 17 and 18 while the effective mass is reduced on average. The angular dependence is also much less smooth at the high temperature, which can be attributed to the more complex Fermi surface topology (Fig.~\ref{Sfig:tempDMFT}(b)). 

